\documentclass[aps,preprint,superscriptaddress,amsmath,amssymb]{revtex4}
\begin{document}
%
\preprint{INFNCA-TH0206}
\title{Super-Kamiokande $hep$ neutrino best fit: a possible signal of
nonmaxwellian solar plasma}
\author{Massimo Coraddu}
\email{massimo.coraddu@ca.infn.it}
\affiliation{Dipart. di Fisica dell'Universit\`a di Cagliari,
             S.P. Sestu Km~1, I-09042 Monserrato (CA), Italy}
\affiliation{Ist. Naz. Fisica Nucleare (I.N.F.N.) Cagliari, 
             S.P. Sestu Km~1, I-09042 Monserrato (CA), Italy}
\author{Marcello Lissia}
\email{marcello.lissia@ca.infn.it}
\affiliation{Ist. Naz. Fisica Nucleare (I.N.F.N.) Cagliari, 
             S.P. Sestu Km~1, I-09042 Monserrato (CA), Italy}
\affiliation{Dipart. di Fisica dell'Universit\`a di Cagliari,
             S.P. Sestu Km~1, I-09042 Monserrato (CA), Italy}

\author{Giuseppe Mezzorani}
\email{giuseppe.mezzorani@ca.infn.it}
\affiliation{Dipart. di Fisica dell'Universit\`a di Cagliari,
             S.P. Sestu Km~1, I-09042 Monserrato (CA), Italy}
\affiliation{Ist. Naz. Fisica Nucleare (I.N.F.N.) Cagliari, 
             S.P. Sestu Km~1, I-09042 Monserrato (CA), Italy} 
\author{Piero Quarati}
\email{quarati@polito.it}
\affiliation{Dipart. di Fisica del Politecnico di Torino, 
             I-10129 Torino, Italy}
\affiliation{Ist. Naz. Fisica Nucleare (I.N.F.N.) Cagliari, 
             S.P. Sestu Km~1, I-09042 Monserrato (CA), Italy}
\date{December 3, 2002}
\begin{abstract}
The Super-Kamiokande best global fit, which includes data from SNO, Gallium
and Chlorine experiments, results in a $hep$ neutrino contribution to the
signals that, even after oscillation, is greater than the SSM prediction.
The solar $hep$ neutrino flux that would yield this contribution is four times
larger than the one predicted by the SSM. Recent detailed calculations exclude 
that the astrophysical factor $S_{hep}(0)$ could be wrong by such a
large factor. Given the reliability
of the temperature and densities profiles inside the Sun, this experimental
result indicates that plasma effects are important for this reaction.
We show that a slight enhancement of the high-energy tail, enhancement that
is of the order of the deviations from the Maxwell-Boltzmann distribution
expected in the solar core plasma, produces an increment of the $hep$ rate
of the magnitude required. We verified
that the other neutrino fluxes remain compatible with experimental signals
and SSM predictions. Better measurements of the high-energy tail of the
neutrino spectrum would improve our understanding of reaction rates in
the solar plasma.
\end{abstract}
\maketitle
\section{\label{sec:intro}Introduction}
Results of the SK and SNO experiments~\cite{Fukuda:2001nk,Ahmad:2001an}
provide model-independent evidence of electon-neutrino oscillation into
other flavors.
A recent global fit of the Super-Kamiokande (SK)
Collaboration~\cite{Fukuda:2002pe}
that includes data of SNO, Gallex/GNO and SAGE shows that the Large
Mixing Angle (LMA) solution is preferred at the 98,9\% confidence level.
The corresponding mass-square difference $\Delta m^2 $ is within the range
$3\cdot 10^{-5} \text{eV}^2 \leq \Delta m^2 \leq 19\cdot 10^{-5} \text{eV}^2$,
and the mixing angle $\theta$ within the range 
$0.25 \leq  \tan^2\theta \leq 0.65$.

The $^8$B flux resulting from the SK best fit,
$(5.33 \pm 0.36 ) \cdot 10^6$~cm$^{-2}\text{s}^{-1}$, is in
substantial agreement with the SSM value~\cite{Bahcall:2000nu},
$(5.05^{+1.01}_{-0.81} ) \cdot 10^6$~cm$^{-2}\text{s}^{-1}$.
Also the $^8$B flux measured by SNO~\cite{Ahmad:2002jz}
assuming the standard $^8$B energy spectrum,
$(5.09^{+ 0.44}_{-0.43}) \cdot 10^6$~cm$^{-2}\text{s}^{-1}$, or
the one obtained using a distorted spectrum,
$(6.42 \pm 1.57) \cdot 10^6$~cm$^{-2}\text{s}^{-1}$,
are in agreement with the SSM value and with the SK fit.
On the contrary, the $hep$ flux resulting from the SK best fit,
\[
\Phi_{hep}^{\text{SK}} = 36 \cdot 10^3 \text{cm}^{-2} \text{s}^{-1} \quad ,
\]
which corresponds to the LMA solution,
is about four times higher than the one predicted in the SSM, 
$9.3 \cdot 10^3$~cm$^{-2}\text{s}^{-1}$~\cite{Bahcall:2000nu},
if one uses $S_{hep}(0)=10.1\cdot 10^{-20}$~keV~barn~\cite{Marcucci:2000bh}, or
\[
\Phi_{hep}^{\text{SSM}} = 7.9 \cdot 10^3 \text{ cm}^{-2} \text{s}^{-1} \quad,
\]
if one uses the more recent 
$S_{hep}(0)=(8.6\pm 1.3)\cdot 10^{-20}$~keV~barn~\cite{Park:2002yp}.
The second best fit (LOW) gives also a $hep$ flux considerable larger than
the one predicted in the SSM, $23 \cdot 10^3 \text{ cm}^{-2} \text{s}^{-1}$.
Only the Quasi-VAC and SMA solutions, which appear disfavored from present
data, give a $hep$ flux compatible with the SSM~\cite{Fukuda:2002pe}.

Given the actual experimental evidence, we should justify a $hep$ flux
four times larger than in the SSM.
Bahcall and Krastev~\cite{Bahcall:1998se}, discussing the possibility that
the $hep$ flux be considerable larger than in the SSM, conclude that they
cannot find a first-principle physical argument demonstrating that the
astrophysical factor $ S_{hep}(0) $ cannot exceed 20 or $40 \cdot 10^{-20} 
\text{ keV barn}$, values which could explain the experimental value of the
$hep$ flux.
However, the latest detailed calculations~\cite{Marcucci:2000bh,Park:2002yp} 
exclude an astrophysical factor of this order of magnitude; the latest
result~\cite{Park:2002yp} gives
\[
    S_{hep}(0) = (8.6 \pm 1.3) \cdot 10^{-20} \text{ keV barn} \quad . 
\]

Excluding that the cross section be wrong by a factor of four, the
average rate can be four times larger only if the product of the thermal
average of the cross section $\langle \sigma v\rangle$ times the $p$
and $^3He$ densities is four times larger. But both the temperature
and the $p$ density are accurately measuread by
helioseismology~\cite{Bahcall:1996qw,Ricci:1997pq}
in the region where $hep$ neutrino are produced. In addition increasing
the temperature or the densities would affect much more the
$^7Be$-, $^8B$- and CNO-neutrino fluxes~\cite{Berezinsky:1995zt}. 
There remains the  possibility of some improbable mechanism that
increases the $^3He$ density, which is not measured by
helioseismology~\cite{Berezinsky:1999pe}, 
by a factor of four in the
$hep$-neutrino production region ($r/R_{\odot} \gtrsim 0.12$) but not in
the $^7Be$-, and $^8B$-neutrino production regions
($r/R_{\odot} < 0.12$). Note that mixing would produce the opposite
effect~\cite{Berezinsky:1999pe}.

We propose that the large increase of $hep$ flux which fits
the present experiemental data is a consequence of the enhancement the
high-energy tail of the $^3He$-$p$ momentum distribution. The
required $hep$ flux of $36 \cdot 10^3 \text{ cm}^{-2} \text{s}^{-1} $
is a signal that the thermal ion distribution of the solar plasma
deviates slightly from the standard one in the region where the reaction 
$^3He  +  p \to ^3He + e^+ +  \nu_e $ is active. In fact small
deviations in the high-energy tail give dramatic effects on the
fusion rates without affecting the bulk properties of the medium, which
are the ones measured by
helioseismology~\cite{Kaniadakis:1997my,Coraddu:1998yb}.

Let us remind a few of the reasons that lead to deviations from the
Maxwellian distribution~\cite{Kaniadakis:1997my,Coraddu:1998yb}.
In plasmas with parameters such as those that
can be found in part of the solar core, in the solar atmosphere or in
the interior of giant planets~\cite{Coraddu:2001ps}, the Debye screening is
approximately valid, but the essential many-body character of the
interaction must be taken into account (nonlocality); the time necessary to
build up again screening after hard collisions is comparable to the
inverse of plasma frequency (memory effect or time nonlocality); 
since many collisions are necessary before particles loose memory of their
initial state, the scattering process cannot be considered Markovian.
The conditions are such that there is no clear separation between
collective and individual degrees of freedom and the description of the
system in the relevant energy range requires the use of additional
scales, dynamically generated by the interaction, in addition to the
temperature: the pure exponential $\exp{(-E/kT)}$, which is determined
by the single scale $kT$, is insufficient.
The Maxwellian regime ends below a given scale.
It has also be shown~\cite{Coraddu:2000nu} that a finite width of
the quasi-particle is effectively equivalent to a distortion of the
momentum distribution.

Small deviations from the Maxwellian distribution, as the ones that are
relevant for us, can be parameterized by the deformation parameter
$\delta$ introduced many years ago by Clayton {\em et al.}~\cite{Clayton:1975}.
This description could be related to the general framework for
non-Maxwellian distributions that has been proposed by Tsallis and applied
in many different contexts~\cite{Tsallis:1988,Latora:ad}: the corresponding
distribution depends on parameter $q$ that can be related to $\delta$ in the
appropriate limit. The parameter $\delta$ can be related to the plasma
parameter $\Gamma$ and the ion-ion correlation  special parameter 
$\alpha$~\cite{Kaniadakis:1997my,Coraddu:1998yb}, and to the finite width
of the quasi-particle~\cite{Coraddu:2000nu}.

In the next Section we derive the $hep$ flux.
In the third Section we present the expression that relates the deformation
parameter $\delta$ or $q$ to the plasma parameter and the ion-ion
correlation  special parameter $\alpha$; in addition we calculate the effect
on the other neutrino fluxes, considering the LMA oscillation and the
admissible deviations from Maxwellian distribution of the different
reacting ions. In the last Section we summarize and comment our results.

\section{\label{sec:hepflux}The $hep$ flux}
The reaction $^3He +  p \to ^4He  + e^+  +  \nu_e $ 
produces a neutrino with an endpoint energy of 18.81~MeV, the
highest energy expected for solar neutrinos and the only one above
15~MeV. The rate of the $hep$ reaction is
very slow and does not affect the solar structure.

The SSM $hep$ flux, which uses standard distribution in the plasma,
is~\cite{Bahcall:1998se}:
\[
\Phi^{\text{SSM}}_{hep} = 2.1 \cdot (1 \pm 0.03)  \cdot
                   \left(\frac{S_{hep}(0)}{2.3\cdot 10^{-20}\text{keV barn}}
                   \right) \cdot
                   10^3 \text{ cm}^{-2} \text{s}^{-1} \quad .
\]
Latest values of $ S_{hep}(0)$ have ranged from 
$(2.3 \pm 0.9)\cdot 10^{-20}$~keV~barn~\cite{Schiavilla:1992} to
$(10.1 \pm 0.9)\cdot 10^{-20}$~\cite{Marcucci:2000bh}.
Let us mention that Alberico {\em et al.}~\cite{Alberico:2000qe} have
proposed a method to determine $S_{hep}(0) $ from a (challanging)
laboratory experiment on electron scattering. 
The latest and more reliable determination~\cite{Park:2002yp} 
\[
    S_{hep}(0) = (8.6 \pm 1.3) \cdot 10^{-20}   \text{keV barn} 
\]
gives
\[
\Phi_{hep}^{\text{th}} 
       = (7.9 \pm 1.4)\cdot  10^3 \text{ cm}^{-2} \text{s}^{-1} \quad ,
\]
where the quoted error includes only the contribution due to $S_{hep}(0)$.

Super-Kamiokande best global fit (LMA) yields a $hep$ flux,
$36 \cdot 10^3$~cm$^{-2}$~s$^{-1}$, about four times 
$\Phi_{hep}^{\text{best}}$, while their 
second best fit (only 1\% of probability) yields 
$23 \cdot 10^3$~cm$^{-2}$~s$^{-1}$, about three times
$\Phi_{hep}^{\text{best}}$~\cite{Fukuda:2001nk}; therefore
\[
  \Phi_{hep}^{\text{exp}} = (3 \div 4)\times \Phi_{hep}^{\text{th}} \quad .
\]

Fusion rates are proportional to the local densities of incoming particles
and to  the average cross section
$\left< v \sigma(v) \right>$
where $v$ is the relative velocity. Therefore, rates
depend on the statistical distribution, {\em i.e.}, on the local
temperature if the distribution is Maxwellian, and also on
at least an additional deformation parameter if the  distribution 
deviates from the Maxwellian. For example, small deviations can be
parameterized by the form proposed by Clayton~\cite{Clayton:1975}:
\begin{equation}
\label{eq:clay}
e^{-E/kT} \to e^{-E/kT-\delta (E/kT)^2} 
\end{equation}
for $\delta>0$; this distribution (Druyvenstein distribution) introduces
a scale $kT/\delta$ in addition to $kT$: the second term in the exponential
becomes important when $E\gtrsim kT/\delta$.

To first approximation, we disregard the changes of the other rates
and write the change of the average $hep$ cross section,
and therefore of the rate, as~\cite{Kaniadakis:1997my,Coraddu:1998yb}:
\begin{equation}
\label{eq:ratechange}
   \left< v \sigma (v) \right>_{\delta_{1,3}} = 
   \left< v \sigma (v) \right>_0  e^{-\gamma_{1,3} \delta_{1,3} } 
\end{equation}
where $\gamma_{1,3} = \left(E_{0} / kT \right)^2_{hep}$ with
$E_0  =  \left( E_G  ( k T )^2 /4 \right)^ {1/3} $; 
$E_G=2\mu c^2(Z_1 Z_2 \alpha \pi)^2$ 
is the Gamow energy and  $\left< v \sigma (v) \right>_0 $ the
Maxwellian rate. The indices $(1,3)$ indicate the reaction $^1$H + $^3$He.
For this reaction $\gamma_{1,3} = 
\left(9.07~\text{keV} / 1.01~\text{keV} \right)^2 \approx 81$.
For small enhancement of the tail of the distribution the
deformation parameter $\delta$ is negative and related to Tsallis entropic
parameter $q$, $\delta = (1-q )/2 $ with $q>1$. Note that even if the 
distribution of Eq.~(\ref{eq:clay}) is normalized only for $\delta>0$,
the resulting change of the rate in Eq.~(\ref{eq:ratechange}), which is
derived as an asymptotic expansion in $\delta$, can be
analytically continued to $\delta<0$.

The required enhancement
\[
    e^{- \gamma_{1,3} \delta_{1,3} }  = 3 \div 4 
\quad\quad  \text{ implies} \quad\quad
     -0.017 \lesssim \delta_{1,3} \lesssim  -0.014 \quad . 
\]

Let us compare this range of values of $\delta$ with
the relation derived and discussed in
Refs.~\cite{Kaniadakis:1997my,Coraddu:1998yb}
\[
\left| \delta_{i,j} \right| = 12 \alpha^4_{i,j} \Gamma_{i} \Gamma_{j}\quad ,
\]
which gives $\delta_{i,j}$ in terms of
$\Gamma_i=Z_i\sum_j Z_j e^2 (n_j)^{1/3}/kT$,
the plasma (or ion-ion) coupling constants
($\Gamma_i\approx 0.14 Z_i$ in the Sun core),
where $n_j$ is the number density of ion $j$, 
and of $\alpha_{i,j}$, the ion-sphere-model parameters introduced by
Ichimaru~\cite{Ichimaru:1992,Yan:1986}, parameters that are related to the
ion-ion correlation fuctions ($0.4<\alpha<0.9$). An alternative
formula by Ichimaru~\cite{Ichimaru:1993}, which uses
$\Gamma^2_{i,j} \equiv \left( Z_i Z_j e^2 
            \left((n_i^{1/3}+ n_j^{1/3}\right)/kT)\right)^2$ instead of
$ \Gamma_{i} \Gamma_{j} $, yields similar ion-sphere-model paramenters.

\section{\label{sec:SolarNuflux}The other fluxes}
In this Section we want to give a quantitative estimate of the range
of $\delta$'s compatible with the experimental determination of the main
neutrino fluxes, and compare it with the range of values suggested for the 
$hep$ neutrino flux.

Previous calculations that assumed standard neutrinos have shown that
$\delta$  of the order of a few times $10^{-3}$ were compatible with the
neutrino experiments even if these deviations from the Maxwellian distribution
were not sufficient to solve the solar netrino
problem~\cite{Kaniadakis:1997my,Coraddu:1998yb,Coraddu:2000nu}.
Since we now have experimental evidence for oscillations and the best fit
to the oscillation parameters (LMA) yields fluxes close to the one predicted
in the SSM, we expect that the values of $\delta$ compatible with present
data be even closer to zero. 
Therefore,
the effect on the neutrino fluxes of small deviations from the Maxwellian
distribution can be reliably obtained by using power laws that include the
reajustment of the solar model~\cite{Castellani:1996cm,Coraddu:1998yb}.

We repeat the analysis of Ref.~\cite{Coraddu:1998yb} taking into account
the oscillation probability. We add the survival probability factors
$P_i$ of $i$-th neutrino flux to the four equations of Castellani 
{\em et al.}~\cite{Castellani:1996cm}. The Gallium~\cite{Strumia:2002rv},
Chlorine~\cite{Cleveland:nv} and Luminosity equations yield:
\begin{subequations}
\begin{eqnarray}
 70.8 \pm 4.5  & = &  73.185 \cdot R_{pp}  P_{pp}  
                    + 34.196 \cdot R_{Be} P_{Be}
                    +  9.0165 \cdot R_{CNO} P_{CNO} 
                    +  2.43 \cdot \Phi^{SNO}_{B} \\ 
  2.56\pm 0.23 & = &   0.0226 \cdot R_{pp} P_{pp}
                     + 1.1448 \cdot R_{Be} P_{Be}
                     + 0.4239 \cdot R_{CNO} P_{CNO}
                     + 1.11 \cdot \Phi^{SNO}_{B} \\
 63.85         & = &   0.980\cdot 59.85 \cdot R_{pp} 
                     + 0.939 \cdot 4.77 \cdot R_{Be} \nonumber \\
                && + 0.937\cdot 1.034 \cdot R_{CNO} 
                   + 0.498 \cdot 10^{-3} \cdot 5.05 \cdot R_{B}
                \quad ,
\end{eqnarray}
where $5.05$ is the produced $^8$B flux; the
$^8$B flux in units of $10^6$~cm$^{-2}$~s$^{-1}$
\begin{equation}
 \Phi^{SNO}_{B} = (5.05^{ + 1.01}_{ - 0.81})\cdot R_B P_B 
                = 1.76  \pm 0.05
\end{equation}
\end{subequations}
is measured in the experiment SNO. The SSM fluxes $\Phi$ are the ones
calculated by Bahcall {\em et al.}~\cite{Bahcall:2000nu}.
We use the survival probability $P_i$ extracted from the experiements by 
Berezinsky and Lissia~\cite{Berezinsky:2001uv}:
$P_{pp} = 0.58$  at 0.265~MeV, $ P_{Be} = P_{CNO} = 0.55$ at 0.814~MeV,
and  $P_{B}  = 0.32$ at 6.71~MeV.
The quantity $R_i$, defined in Ref.~\cite{Kaniadakis:1997my},
represents the rate enhancement or depletion due to the nonmaxwellian tail.
Eliminating $R_{pp}$ and using the relation between $R$ and
$\delta$~\cite{Kaniadakis:1997my,Coraddu:1998yb}, the other $R_i$ (with
conservative errors) are:
\begin{equation}
 \begin{array}{l c c c r}
 R_{Be}  & = &  e^{23  \delta_{pp} - 139.75 \delta_{He}} & = & 0.87\pm 0.09 \\
 R_{CNO} & = &  e^{62.1\delta_{pp} +    6.5 \delta_{He} - 407\delta_{p}} 
                                                         & = & 0.87\pm 0.09\\ 
 R_{B}   & = &  e^{62.1\delta_{pp} + 149.5  \delta_{He} - 190 \delta_{p}} 
                                                         & = & 1.10 \pm 0.20
 \end{array}
\end{equation}
where for the purpose of this estimation we have introduced several
$\delta$'s: $\delta_{pp}$ for protons in the $pp$ production
region ($r/R_{\odot} \gtrsim 1$), which is also the region where the 
$hep$ neutrinos are produced, $\delta_p$ for protons in the inner core
($r/R_{\odot} \lesssim 1$), where all other neutrinos are produced, and
$\delta_{He}$ for the distribution of the $He$ nuclei ($^3$H and $^4$H).
The result is that $\delta_{pp} \sim - 10^{-2}$, while
$|\delta_p | \sim |\delta_{He}|\lesssim 2\cdot 10^{-3}$.

We note that neutrinos from the $pp$ and $hep$ reactions are produced in
the same region ($r/R_{\odot} \gtrsim 1$), and they have similar values 
(limit) for $\delta$. The difference is that values of $\delta \sim 2\%$
have large effect on the $hep$ reaction~\cite{Coraddu:2000nu}, while they
have a much smaller effect on the $pp$ reaction
both because the Coulomb barrier is lower and because the adjustment of the
solar structure to a change of the $pp$ rate, which is directly linked to the
luminosity, is such that it minimizes the change of the rate, that ends up
being between  0.5\% and 3.0\% larger that in the SSM (at the solar surface).

The other neutrino fluxes are all produced in the
inner core ($r/R_{\odot} \lesssim 1$), where the plasma conditions are
different, and they show much smaller deviation from Maxwellian
distribution. We note that the value of $\alpha$ deduced from the
$\delta$ of the proton component, {\em i.e}, a value between 0.54 and 0.72,
is in agreement with the range of values allowed for a weakly coupled and
weakly non ideal plasma ($\Gamma \lesssim 1 $). In addition 
the values of the different parameters $\delta$ are within the
constraints imposed by helioseismology, as it was checked in
Ref.~\cite{Degl'Innocenti:dy}  without considering oscillations: oscillations
makes the agreement even better. In detail protons superdiffusion in the
inner core would yield a $^7$Be flux reduced by 4--22\% with 
respect to SSM calculations; the same range of reduction is
obtained for the CNO flux (these fluxes are at the solar surface, before
oscillation).

Such values of $\delta$ could be measured or excluded by more
precise measurements and solar model calculations. 

\section{\label{sec:Conclusion}Discussion and conclusion}
Our work is based on the following points:
\begin{itemize}
\item
the global best fit (LMA) to the neutrino experiments requires a $hep$-neutrino
produced flux of $36 \cdot 10^3$~cm$^{-2}$~s$^{-1}$~\cite{Fukuda:2001nk},
about four times the one predicted by the SSM~\cite{Bahcall:2000nu}; also
the second best fit (LOW) requires a flux three times the one in the SSM;
if one consider the contribution to the signals on earth from $hep$ neutrinos,
this contribution, even after oscillation is larger than the SSM prediction;
\item
the latest precise evaluation~\cite{Park:2002yp} of the astrophysical factor,
$ S_{hep}(0) = (8.6 \pm 1.3 ) \cdot 10^{-20} $~keV~barn, which is
slighter smaller than the values used in the SSM~\cite{Marcucci:2000bh},
rules out that the cross section could be a factor of four larger;
\item
temperature and $p$ density are measured very precisely by helioseismology
in the region of interest and could not vary significantly without
dramatically affecting the $^8$B- and $^7$Be-neutrino fluxes;
\item
it is not known a physical mechanism that could increase the density of
$^3$He in the $hep$-neutrino production region without increasing the
same density in the $^8$B- and $^7$Be-neutrino production region (any
mixing would produce the opposite effect)~\cite{Berezinsky:1999pe};
\item
there exist several mechanisms that deform the high-energy
tail of the ion velocity distribution in 
plasmas~\cite{Kaniadakis:1997my,Coraddu:1998yb,Lavagno:2001nm,Coraddu:2000nu};
we do not exclude that other mechanisms could be found that also produce
deformation of the distribution.
\end{itemize}

We have proposed that the experimental results are signals of plasma
effects on the fusion rates.

We have verified that the slight enhancement of
the high-energy tail of the distribution
naturally produces the needed large $hep$-neutrino flux; the resulting
deformation parameter is in agreement with the theory of weakly coupled
and weakly non-ideal plasmas.
This kind of modifications of the distribution do not affect the
bulk properties and, in particular, the measured helioseismiologic quantities.
The other neutrino fluxes are not significantly modified.

Future more precise experimental determinations of the oscillation
parameters, $\Delta m^2 $ and mixing angle $\theta$,
and of the high-energy neutrino spectrum (and in particular
a direct measurement of the $hep$ contribution)
and improved theoretical calculations, or experimental
determination~\cite{Alberico:2000qe}, of the astrophysical factor
$S_{hep}(0)$ could corroborate (or bound) the importance of plasma
effects for some of the fusion rates in the Sun.  

\begin{acknowledgments}
This work is partially supported by M.I.U.R. (Ministero dell'Istruzione,
dell'Universit\`a e della Ricerca) under Cofinanziamento P.R.I.N. 2001.
\end{acknowledgments}

\end{document}